\documentclass[12pt]{iopart}

\usepackage{graphicx}
\begin{document}
\title{Quantum entanglement of a harmonic oscillator in an electromagnetic field}
\author{D.\,N. Makarov}

\address{Federal Center for Integrated Arctic Research, Russian Academy of Sciences, nab. Severnoi Dviny 23, 163000, Arkhangelsk, Russia.}

\address{Northern (Arctic) Federal University, nab. Severnoi Dviny 17, 163002, Arkhangelsk, Russia.}
\ead{makarovd0608@yandex.ru}
\vspace{10pt}
\begin{indented}
\item[]September 2017
\end{indented}
\begin{abstract}
At present, there are many methods of quantum entanglement of particles with an electromagnetic field. Most methods have a low probability of quantum entanglement and not an exact theoretical apparatus based on an approximate solution of the Schrodinger equation. There is a need for new methods for obtaining quantum-entangled particles and mathematically accurate studies of such methods. In this paper, a quantum harmonic oscillator (for example, an electron in a magnetic field) interacting with a quantized electromagnetic field is considered. Based on the exact solution of the Schrodinger equation for this system, it is shown that for certain parameters there can be a large quantum entanglement between the electron and the electromagnetic field. Quantum entanglement is analyzed on the basis of a mathematically exact expression for the Schmidt modes and the Von Neumann entropy. 
\end{abstract}

\section{Introduction}

It is known that quantum entanglement arises when the wave function of a particle system cannot be represented as a product of the wave functions of each particle. There are many such systems from which to choose. The basic method of quantum entanglement of particles is spontaneous parametric down-conversion (SPDC) \cite{Burnham_1970, Kwiat_1995, Fulconis_2007}. In this process, when a nonlinear birefringent crystal is irradiated by a laser pump field with a certain low probability, the pump photons decay - with a low probability - into two quantum-entangled photons of lower frequencies. Despite the low probability of the decay of the pump photon, SPDC has been widely studied and has many practical applications in quantum computer science. This is due to the fact that this method is the simplest for obtaining quantum-entangled photons. While other more sophisticated methods of generating entangling particles have been proposed \cite{Aspect_1981}-\cite{Muller_2009} none of them result in a greater probability of entanglement of the particles under consideration than SPDC \cite {Dousse_2010, Korzh_2015}. There is also a difficulty with the theoretical study of the various methods, which is that basically, quantum entanglement is a complex process from the mathematical point of view. For the systems under consideration, it is impossible to accurately study the quantum dynamics of the models under consideration; it is impossible to solve exactly the nonstationary Schrodinger equation and calculate the entanglement parameter (for example, the Von Neumann entropy in the case of a two-component system). Solutions are mainly sought using higher orders of perturbation theory and on the basis of these solutions conclusions are drawn about quantum entanglement. It is known that perturbation theory is applicable if there is a weak interaction of particles in the system, and therefore use of this theory is based on the assumption that quantum entanglement is not a large quantity. In order to study a system with large quantum entanglement, the exact solution of the Schrodinger equation must be used - a task that is very difficult to achieve mathematically. Nonetheless, when studying quantum entanglement, such a solution would be of fundamental interest. In quantum physics, it is known that exact solutions of the Schrodinger equation do not present any great difficulty. In studying quantum entanglement, one can include a two-level model of Jaynes-Cummings \cite{Jaynes_Cummings} in order to achieve sufficiently accurate solutions to mathematical methods. However, it should be noted that this model is not an exact solution of the nonstationary Schrodinger equation. Also some model problems and particular cases for the stationary Schrodinger equation \cite {Han_Kim_1999, Peschel_2009}. Quantum Dynamics of quantum-entangled systems is being studied actively at the present time \cite{Plenio_2004}-\cite{Makarov_2017_adf}. For example, on the basis of the exact solution of the nonstationary Schrodinger equation for an atom in a strong two-mode electromagnetic field, it has been shown \cite {Makarov_2017_adf, Makarov_2017_TMF} that there is a strong quantum entanglement between photons. However, as this solution is based upon the assumption that the external electromagnetic field is many times stronger than the Coulomb field of attraction of an electron in an atom, it does not promote an understanding of the influence of the Coulomb field on quantum entanglement. Therefore, the problem of finding a system whose quantum entanglement is large, and also in which it would be possible to mathematically study quantum entanglement on the basis of an exact solution of the Schrodinger equation, is topical. In addition to this relevance, the problem of practical application is important, i.e. the proposed method should be simple enough for practical implementation.

In this paper we study the method of quantum entanglement of a harmonic oscillator with an external electromagnetic field. A harmonic oscillator can be an electron in a uniform magnetic field with induction $ \bf B $. It is known \cite {Landau_3} that in such a field the electron behaves like a harmonic oscillator with frequency $ \omega_ {c} = \frac {B e} {m_ {e} c} $, where $ c $ is the speed of light, $ m_ {e} $ and - $ e $ are, the mass and charge of an electron, respectively. The wave function of such a system can be found on the basis of the exact solution of the Schrodinger equation of the specific system being studied. The quantum entanglement of the system can be studied using the analytical form of the Schmidt mode \cite {Grobe_1994, Ekert_1995} and the von Neumann entropy \cite {Bennett_1996, Casini_2009}. It is shown that for certain parameters of the system under consideration, a large quantum entanglement is possible.

\section{ Exact solution of the Schrodinger equation}

Let us consider, at a time $t>0$, a quantized electromagnetic field acting on a system consisting of a quantum harmonic oscillator. The Schrodinger equation for this system will be
\begin{equation}
i\hbar \frac{\partial \Psi}{\partial t}=\left\lbrace \frac{1}{2 m_{e}}\left( -i\hbar \frac{\partial }{\partial {\bf {r}}}+\frac{e}{c}{\hat{\bf{ A}}}\right)^2 + {\hat H}_{f} + \frac{\omega^2_{c}(x^2+\delta (y^2+z^2))}{2} \right\rbrace \Psi.
\label{1}
\end{equation}
In the expression: (\ref{1}) $\delta =0$ corresponds to one type of interaction, and $\delta =1$ to another type; $i$ is the imaginary unit; ${\hat H}_{f}=\sum_{\bf {k},\bf {u}}\hbar \omega \left( {\hat a}^{+}_{\bf {k},\bf {u}} {\hat a}_{\bf {k},\bf {u}}+\frac{1}{2}\right) $ is the Hamiltonian of the electromagnetic field, where ${\hat a}^{+}_{\bf {k},\bf {u}}$ and ${\hat a}_{\bf {k},\bf {u}}$  are the creation and annihilation operators, respectively, for photons with the wave vector ${\bf k}$ and the polarization $\bf {u}$; and the vector potential of the electromagnetic field ${\hat{\bf{ A}}}$ has the form
\begin{equation}
{\hat{\bf{ A}}}=\sum_{\bf {k},\bf {u}} \sqrt{\frac{2\pi c^2 \hbar }{\omega V_{f}}}\left({\bf {u}}exp\left(i(\omega t - {\bf k r}) \right){\hat a}_{\bf {k},\bf {u}} + {\bf {u}}^*exp\left(-i(\omega t - {\bf k r}) \right){\hat a}^{+}_{\bf {k},\bf {u}} \right) .
\label{2}
\end{equation}
The summation for ${\hat H}_{f}$ and expression (\ref{2}) include all possible values of the wave vector ${\bf k}$ and the polarization $\bf {u}$. 
Next we consider a single-mode field, in which $\sum_{\bf {k},\bf {u}} $ has one summand with the wave vector ${\bf k}$ with polarization ${\bf u}$.
Further, for convenience, we apply the atomic system of units $\hbar=1, m_{e}=1, e =1$. Now consider expression (\ref{1}) in the dipole approximation in which the vector potential, expressed in terms of field variables, is equal to ${\hat{\bf{ A}}}=a{\bf u} q$, where $a=\sqrt{\frac{4\pi c^2}{\omega V}}$, $\omega$ is the frequency of the mode under consideration, $V$ is the quantization volume, and $q $ is the field variable of the mode. Since we are considering a single-mode electromagnetic field, it is convenient to orient the polarization vector along the $x$ axis i.e. ${\bf u}={\bf i}$, where ${\bf i}$ is the unit vector  directed along the $ x $ axis (the same assumption applies throughout the paper).

For example, expression (\ref{1}) for $\delta =0$ corresponds to the case of an electron in a magnetic field interacting with an external quantized electromagnetic field if the polarization of the electromagnetic field $\bf {u}$ is in the plane perpendicular to the magnetic induction vector $\bf B$. This is easily obtained by using the Landau gauge \cite{Landau_3} for the magnetic field ${\bf A}_{L}=B x {\bf j}$, where ${\bf j}$ is the unit vector directed along the $y$ axis.

As a result, the dynamics of the system, and electronic transitions are not determined by the stationary Schrodinger equation with the Hamiltonian 
\begin{equation}
{\hat{H}}= \frac{1}{2}\left( -i \frac{\partial }{\partial x}+\beta q \right)^2 + \frac{\omega}{2}\left(q^2-\frac{\partial^2 }{\partial {q^2}}\right)+\frac{\omega^2_{c}x^2}{2} ,
\label{3}
\end{equation}
where $\beta=\sqrt{\frac{4\pi}{\omega V}}$. It should be clarified that expression (\ref{3}) corresponds to the Hamiltonian without the variables $y,z$, since the wave function corresponding to the solution of equation (\ref{1}) $\Psi (x,y,z,q,t)=\Psi (y,z,t)\Psi (x,q,t)$, and this means that $\Psi (y,z,t)$ is not a perturbed wave function and therefore does not affect the electronic transitions of the system under the action of an external electromagnetic field. Extending the argument, we replace $x\to \sqrt{\omega_{c}}x$, and the Hamiltonian (\ref{3}) becomes
\begin{equation}
{\hat{H}}= \frac{\omega}{2}\left(q^2-\frac{\partial^2 }{\partial {q^2}}\right)+\frac{\omega_{c}}{2}\left(x^2-\frac{\partial^2 }{\partial {x^2}}\right)+\frac{\beta^2}{2}q^2-i\beta \sqrt{\omega_{c}}q \frac{\partial }{\partial x} .
\label{4}
\end{equation}
We consider the nonstationary Schrodinger equation with the Hamiltonian (\ref{4}). We represent the differential equation in question in the form
\begin{eqnarray}
{\hat{H}}^{'} \Psi^{'}=i\hbar \frac{\partial \Psi^{'}}{\partial t} ,
\label{5}
\end{eqnarray}
where ${\hat{H}}^{'}={\hat{S}}{\hat{H}}{\hat{S}}^{-1}$ and $\Psi^{'} = {\hat{S}}\Psi$, with ${\hat{S}}$ being a unitary operator and ${\hat{S}}^{-1} $ being the inverse operator of ${\hat{S}}$ (i.e. ${\hat{S}}^{-1}{\hat{S}}=1 $). Finding the solution of the Schrodinger equation (\ref{5}), we obtain the required wave function using the inverse transformation $\Psi = {\hat{S}}^{-1}\Psi^{'}$. We seek a solution for $\Psi^{'}$ in the form, where the operator 
\begin{eqnarray}
{\hat{S}}=exp\left( i \gamma \frac{\partial }{\partial x}\frac{\partial }{\partial q} \right) exp\left(i\alpha q x \right) .
\label{6}
\end{eqnarray}
It should be noted that the choice of the operator ${\hat{S}}$ in expression (\ref{6}) is determined as a result of a careful analysis of the stationary Schrodinger equation with the Hamiltonian (\ref{4}) for the possibility of its diagonalization. In expression (\ref{6}), the constant values $\gamma, \alpha $ are unknown: our task now is to find values of  $\gamma, \alpha $ so that the Hamiltonian ${\hat{H}}^{'}$ in (\ref{5}) becomes diagonal.

Knowing the form of the operator $ \hat{S}$ in expression (\ref{6}) and using known properties of the operators ($e^{\hat{A}}{\hat{H}}e^{-\hat{A}}=\hat{H}+[{\hat{A}},{\hat{H}}]+1/2[{\hat{A}},[{\hat{A}},{\hat{H}}]]+...$ ) it is not difficult to obtain the diagonal Hamiltonian ${\hat{H}}^{'}$; all that is required is to find the appropriate parameters $\gamma, \alpha $. As a result, we get
\begin{eqnarray}
\alpha = \sqrt{\frac{\omega_c}{\omega}}\left(\epsilon \mp \sqrt{\epsilon^2+1} \right),~ \gamma =\pm \frac{1}{2}\sqrt{\frac{\omega}{\omega_c}}\frac{1}{\sqrt{\epsilon^2+1}},~ \epsilon =\frac{\omega^2-\omega^2_c +\beta^2 \omega}{2\beta \sqrt{\omega}\omega_c}.
\label{7}
\end{eqnarray}
When $\beta \to 0$,the system must go to the initial state, for this it is necessary to use the upper sign in expression (\ref{7}) (and elsewhere below) for $\epsilon > 0$,and the lower sign when $\epsilon < 0$. The case with $\epsilon =0$ will be considered below. As a result, the expression for ${\hat{H}}^{'}$ takes the form
\begin{eqnarray}
{\hat{H}}^{'}=\frac{\omega}{2}\left(\Lambda q^2-\sigma \frac{\partial^2 }{\partial q^2} \right) + \frac{\omega_c}{2}\left(\frac{1}{\sigma} x^2-\kappa \frac{\partial^2 }{\partial x^2} \right), 
\label{8}
\end{eqnarray}
\begin{eqnarray}
\Lambda = 1 + \alpha^2\frac{\omega_c}{\omega} -\frac{2\beta \alpha \sqrt{\omega_c}}{\omega}+\frac{\beta^2}{\omega},~~\sigma =\frac{1}{1+\alpha^2\frac{\omega}{\omega_c} },
\nonumber\\ 
\kappa = \sigma +\frac{2\beta \epsilon \gamma^2}{\sqrt{\omega}}-\frac{2\beta \gamma}{\sqrt{\omega_c}}\left(1+\alpha \gamma \right) .
\label{9}
\end{eqnarray}
It is not difficult to solve the Schrodinger equation with the Hamiltonian (\ref{8}), since all the variables are separated. The solution for the nonstationary equation will be sought in the form $\Psi^{'}(x,q,t)=\sum_{n,m}A_{n,m}e^{-i{E_{n,m}}t}\Psi_{n,m}^{'}(x,q)$, where ${\hat{H^{'}}}\Psi_{n,m}^{'}(x,q)=E_{n,m}\Psi_{n,m}^{'}(x,q)$, $n,m = 0,1,2,...$ are quantum numbers, $A_{n,m}$ are coefficients that depend on the initial conditions, and $E_{n,m}$ are the eigenvalues of the energy. First, we write the eigenvalue of the energy of the Hamiltonian (\ref{8})
\begin{eqnarray}
E_{n,m}=\omega_c \left( n+\frac{1}{2}\right) \sqrt{G}+\omega\left( m+\frac{1}{2}\right) \sqrt{S}~ ,
\label{10}
\end{eqnarray}
where
\begin{eqnarray}
G=1 \mp \frac{\beta \sqrt{\omega}}{2\omega_c \sigma \sqrt{\epsilon^2+1}}~,~~ S=1- \frac{\beta \omega_c}{\omega^{3/2}}\left(\epsilon \mp \sqrt{\epsilon^2+1} \right) +\frac{\beta^2}{\omega} .
\label{11}
\end{eqnarray}
Next, we write out the wave function of the Hamiltonian (\ref{8}) in the form $\Psi^{'}_{n,m}(x,q)=\Psi^{'}_{n}(x)\Psi^{'}_{m}(q)$, 
where
\begin{eqnarray}
\Psi^{'}_{n}(x)= C_{n}e^{-\frac{R}{2}x^2}H_{n}\left(x \sqrt{R} \right)~,~~ \Psi^{'}_{m}(q)=C_{m}e^{-\frac{s}{2}q^2}H_{m}\left(q s \right),
\label{12}
\end{eqnarray}
and where $H_{n},H_{m}$ are Hermite polynomials, $R=\sqrt{\frac{1}{\sigma \kappa}}$, $s=\sqrt{\frac{\Lambda}{\sigma}}$, and the normalization coefficients
\begin{eqnarray}
C_{n}=\frac{1}{\sqrt{2^n n!\sqrt{\pi}}} R^{1/4} ,~~~~~~ C_{m}=\frac{1}{\sqrt{2^m m!\sqrt{\pi}}} s^{1/4}.
\label{13}
\end{eqnarray}
We then find $\Psi(x,q,t)={\hat{S}}^{-1}\Psi^{'}(x,q,t)$. As a result, $\Psi(x,q,t)=\sum_{n,m}A_{n,m}e^{-i{E_{n,m}}t}\Psi_{n,m}(x,q)$, where the wave function $\Psi_{n,m}(x,q) = {\hat{S}}^{-1}\Psi^{'}_{n,m}(x,q)$. The action of the operator  ${\hat{S}}^{-1}$ on the function $\Psi^{'}_{n,m}(x,q)$  is not an obvious problem. For this, for example, the wave function $\Psi^{'}_{n}(x)$ must be represented in the form of a Fourier integral $\Psi^{'}_{n}(x)=\frac{1}{\sqrt{2\pi}}\int^{\infty}_{-\infty}a_n(p)exp(ip\sqrt{\frac{1}{\sigma \kappa}}x)dp$, where for the inverse Fourier transform we obtain $a_n(p)=C_n(-i)^n exp(-p^2/2)$ and only then and only then can the operator ${\hat{S}}^{-1}$ act. As a result, we get
\begin{eqnarray}
\Psi_{n,m}(x,q)=\frac{C_n C_m (-i)^n}{\sqrt{2\pi}}\int^{\infty}_{-\infty}e^{-\frac{p^2}{2}} H_n(p)e^{ix\left( p\sqrt{R} -\alpha q\right)}e^{-\frac{s}{2}\left( q+p \gamma \sqrt{R}\right)^2 }\times
\nonumber\\ 
\times H_m\left( \sqrt{s}\left( q+p \gamma \sqrt{R} \right) \right) d p .
\label{14}
\end{eqnarray}
Function (\ref{14}) can be calculated analytically using the table integral \cite{Prudnikov_T3}, but the result is complicated enough without doing further analysis. It will be shown later that for calculating probabilities and quantum entanglement, the analytic form of the function (\ref{14}) is not required, so this calculation is not given here.

Consider the Fourier integral for the wave function $\Psi(x,q,t)$ in the form $\Psi(p,q,t)=\frac{1}{\sqrt{2\pi}}\int^{\infty}_{-\infty}\Psi(x,q,t)exp\left( i p x\right) d x$. After a simple calculation we find that $\Psi(p,q,t)=\sum_{n,m}A_{n,m}e^{-i{E_{n,m}}t}\Psi_{n,m}(p,q)$, where
\begin{eqnarray}
\Psi_{n,m}(p,q)=c_n c_m i^n\left(\frac{s}{R} \right)^{1/4} e^{\frac{1}{2R}\left(p-\alpha q \right)^2 }H_n\left( \frac{1}{\sqrt{R}}\left(p-\alpha q \right) \right)\times
\nonumber\\ 
\times e^{-\frac{s}{2}\left(q(1+\alpha\gamma) -\gamma p\right)^2 } H_m\left(\sqrt{s}\left(q(1+\alpha\gamma) -\gamma p\right) \right),
\label{15}
\end{eqnarray}
and where $c_{n}=\left( \sqrt{2^n n!\sqrt{\pi}}\right)^{-1} ,~ c_{m}=\left( \sqrt{2^m m!\sqrt{\pi}}\right)^{-1} $. We show that the Fourier transform of the wave function can also be used, like the wave function itself, to calculate the probabilities. We expand the wave function in the eigenfunctions of the unperturbed system in the form $\Psi(x,q,t)=\sum_{m_1, m_2}a_{m_1,m_2}(t)\Phi_{m_1}(x)e^{-iE_{m_{1}}t}\Phi_{m_2}(q)e^{-iE_{m_{2}}t}$, where $E_{m_1},E_{m_2}$ are the energy of states, with quantum numbers  ${m_1, m_2}$ ,respectively, for a noninteracting oscillator in the state $|m_1\rangle $ and a free electromagnetic field in the Fock state  $|m_2\rangle $ and $a_{m_1, m_2}(t)$ is the probability amplitude for finding the system in state $\Phi_{m_1}(x), \Phi_{m_2}(q)$ at time $t$. By the definition of probability $w_{m_1,m_2}=\left| a_{m_1, m_2}(t)\right|^2 $ is the probability of detecting a system in the state $m_1, m_2$ at time $t$. Carrying out the Fourier transform over this decomposition (as shown above), we obtain $\Psi(p,q,t)=i^{m_{1}} \sum_{m_1, m_2}a_{m_1,m_2}(t)e^{-iE_{m_{1}}t}e^{-iE_{m_{2}}t}\Phi_{m_1}(p)\Phi_{m_2}(q)$. Using the orthogonality condition, we obtain $a_{m_1, m_2}(t)=(-i)^{m_1}e^{iE_{m_{1}}t}e^{iE_{m_{2}}t}\left\langle \Phi_{m_1}(p)\Phi_{m_2}(q)|\Psi(p,q,t)\right\rangle =(-i)^{m_1}\sum_{n,m}A_{n,m}e^{-i{\Delta E_{n,m}}t}\left\langle \Phi_{m_1}(p)e^{-iE_{m_{1}}t}\Phi_{m_2}(q)e^{-iE_{m_{2}}t}|\Psi_{n,m}(p,q)\right\rangle $, where $\Delta E_{n,m}=E_{n,m}-E_{m_1}-E_{m_2}$. 

In order to calculate $a_{m_1, m_2}(t)$ must be known $A_{n,m}$; these are sought from the initial conditions of the problem $\Psi(p,q,0)=\frac{1}{\sqrt{2\pi}}\int^{\infty}_{-\infty}\Psi(x,q,0)exp(ip x)dx$. Assume that at the initial instant of time  $t=0$ the system was in the state $\Psi(x,q,0)=\Phi_{s_1}(x)\Phi_{s_2}(q)$. The initial state of the system is a noninteracting oscillator in the state $|s_1\rangle $ and a free electromagnetic field in the Fock state $|s_2\rangle $. As a result, $\Psi(p,q,0)=i^{s_1}\Phi_{s_1}(p)\Phi_{s_2}(q)$ and using the orthogonality properties of $\left\langle \Psi_{n,m}(p,q)|\Psi_{n^{'},m^{'}}(p,q)\right\rangle =\delta_{n,n^{'}}\delta_{m,m^{'}}$ ($\delta_{n,n^{'}}$ is the Kronecker symbol), we obtain
\begin{eqnarray}
A_{n,m}=i^{s_1}\left\langle \Psi_{n,m}(p,q)|\Phi_{s_1}(p)\Phi_{s_2}(q) \right\rangle . 
\label{16}
\end{eqnarray}
If in the expression (\ref{16}) the coefficient $A_{n,m}$ is renamed so that it reflects information about the state from which it is calculated $A_{n,m}\to A^{s_1,s_2}_{n,m}$, we then obtain
\begin{eqnarray}
a_{m_1, m_2}(t)=\sum_{n,m}A^{s_1,s_2}_{n,m}A^{*{m_1,m_2}}_{n,m}e^{-i{\Delta E_{n,m}}t} .
\label{17}
\end{eqnarray}
Next, we make a simplification, based on the fact that the parameter $\beta$ entering expression (\ref{17}) is a small quantity. Indeed, for a realistic microcavity or focal volume \cite{Tey_2008}, this value is of the order of $10^{-5} - 10^{-3}$, although it is usually even less. It can be seen from expressions (\ref{7}),(\ref{9}) and (\ref{15})that the quantum entanglement of the system will be significant if the parameter $\epsilon$ is a finite value. Since $\beta<<1$, and $\epsilon$ is a finite quantity, we get $\omega \approx \omega_{c}\approx \omega$,and $\Delta \omega = \omega-\omega_{c}$ is less than or of order $\beta$.
We write out expression (\ref{15}) retaining the main terms under the condition $\beta<<1$ and $\omega \approx \omega_{c}\approx \omega$
\begin{eqnarray}
\Psi_{n,m}(p,q)=i^n c_{n}c_{m} exp\left( -\frac{\sigma}{2}\left(p-\alpha q\right) ^2\right) \times
\nonumber\\ 
H_{n}\left(\sqrt{\sigma}\left(p-\alpha q\right) \right)exp\left( -\frac{\sigma}{2}\left( q+\alpha p \right)^2\right)
H_{m}\left(\sqrt{\sigma}\left(q+\alpha p \right) \right),
\label{18}
\end{eqnarray}
where $\sigma$ in this case will be $\sigma=\left( 1+\alpha^2\right)^{-1} $, and $\alpha=\epsilon \mp \sqrt{\epsilon^2+1}$. The parameter $\alpha$ is the main parameter of the problem under consideration and $\alpha \in (-1,1)$, and $\sigma \in (1/2,1)$. It should be noted that the wave function (\ref{18}) can be obtained by rotating the orthogonal coordinate system [the coordinates $(p,q)\to (p^{'},q^{'})$, where $p^{'}=p \cos\theta - q \sin \theta $, and $q^{'}=q \cos \theta + p \sin \theta $] by an angle $\theta$ corresponding to the value $tg\theta=\alpha$.
 
In order to calculate (\ref{16}) with the wave function (\ref{18}), the results of previous research \cite{Makarov_2017_adf} must be used where an integral of this kind was calculated. As a result, we obtain
\begin{eqnarray}
A^{s_1,s_2}_{n,m}=\frac{i^{s_1-n}(-1)^{s_2+m}\alpha^{s_{2}+m}\sqrt{n!m!}}{(1+\alpha^2)^{\frac{s_1+s_2}{2}}\sqrt{s_{1}!s_{2}!}}P^{(-(1+s_1+s_2), n-s_2)}_{m}\left(-\frac{2+\alpha^2}{\alpha^2} \right),
\label{19}
\end{eqnarray}
where $P^{(b,c)}_{a}(x)$ is the Jacobi polynomial. In expression (\ref{19}), as was shown in the previous study in calculating the integral \cite{Makarov_2017_adf}, the condition $s_1+s_2=m+n$ is fulfilled. If the condition $s_1+s_2=m+n$ is satisfied, then $m_1+m_2=m+n$ is executed, which means $s_1+s_2=m_1+m_2$. This is a very important condition; as will be shown below, its use permits the calculation of the Schmidt modes.

Thus, the probability amplitude for detecting a system in the state $|m_1 \rangle  |m_2 \rangle  $ during the evolution of the system from the state $|s_1 \rangle  |s_2 \rangle  $ is obtained in the analytical form and is determined by the expressions (\ref{10}), (\ref{19}) and
\begin{eqnarray}
a_{m_1, m_2}(t)=\sum^{s_1+s_2}_{n=0}A^{s_1,s_2}_{n,s_1+s_2-n}A^{*{m_1,m_2}}_{n,s_1+s_2-n}e^{-i{\Delta E_{n,s_{1} + s_{2}-n}}t} .
\label{20}
\end{eqnarray}
In the expression (\ref{20}), up accurate to an inessential phase,  $\Delta E_{n,s_{1} + s_{2}-n}$ for $\beta<<1$ can be replaced by $\Delta E_{n,s_{1} + s_{2}-n}\to \delta n$, where $\delta = \beta \sqrt{\omega}\left( \alpha + \epsilon \right) $.  In addition, the wave function $\Psi(x,q,t)$ is accurate to an inessential phase, and taking into account the fact that $s_1+s_2=m_1+m_2$, it can be represented in the form
\begin{eqnarray}
\Psi(x,q,t)=\sum^{s_1+s_2}_{m_1=0}a_{m_1,s_1+s_2-m_1}(t)\Phi_{m_1}(x)\Phi_{s_1+s_2-m_1}(q) .
\label{21}
\end{eqnarray}

\section{ Schmidt's modes and quantum entanglement}

According to Schmidt's theorem \cite{Grobe_1994, Ekert_1995}, the wave function of the system can be expanded in the form $\Psi = \sum_{k}\sqrt{\lambda_k}u_{k}(x)v_{k}(q) $, where $u_{k}(x)$ is the wave function of the pure state 1 of the system, $v_{k}(p)$ is the wave function of the pure state 2 of the system, and where $\lambda_k$ is the Schmidt mode and is an eigenvalue of the reduced density matrix, i.e. $\rho_1(x,x^{'})=\sum_{k}\lambda_k u_{k}(x)u^{*}_{k}(x^{'}) $ or $\rho_2(q,q^{'})=\sum_{k}\lambda_k v_{k}(q)v^{*}_{k}(q^{'}) $. Having found the Schmidt mode $\lambda_k$, the measure of the quantum entanglement of the system can be calculated. To do this, various measures of entanglement can be used, for example, the Schmidt parameter \cite{Grobe_1994, Ekert_1995} $K=\left(\sum_{k}\lambda^2_k \right)^{-1} $ or Von Neumann entropy \cite{Bennett_1996,Casini_2009} $S_N=-\sum_{k} \lambda_k \ln \left(\lambda_k \right) $. The main difficulty in calculating quantum entanglement is the search for $\lambda_k$ for a quantum-dynamic system.

Using expression (\ref{21}), the definition of the reduced density matrix can be obtained
\begin{eqnarray}
\rho_1(x,x^{'})=\sum^{s_1+s_2}_{m_1=0}|a_{m_1,s_1+s_2-m_1}(t)|^2 \Phi_{m_1}(x)\Phi^{*}_{m_1}(x^{'}),
\nonumber\\ 
\rho_2(q,q^{'})=\sum^{s_1+s_2}_{m_1=0}|a_{m_1,s_1+s_2-m_1}(t)|^2 \Phi_{s_1+s_2-m_1}(q)\Phi^{*}_{s_1+s_2-m_1}(q^{'}) ,
\label{22}
\end{eqnarray}
from which it follows that $\lambda_k=|a_{k,s_1+s_2-k}(t)|^2$, and $S_N=-\sum^{s_1+s_2}_{k=0} \lambda_k \ln \left(\lambda_k \right) $.

We now present the results of quantum entanglement for the Von Neumann entropy $S_N$. As an example, in Figure 1 we consider the dependences of $S_N=S_N(\delta t)$ on the dimensionless parameter for $\alpha = (1,3/4,1/2,1/10,1/100)$ and four variants of photon numbers $s_2$ and oscillator $s_1$. Calculation of entropy, for negative values of  $\alpha$, is not necessary, since $\lambda_k$ is an even function $\lambda_k(\alpha)=\lambda_k(-\alpha)$. From this it follows that $\lambda_k$, and hence the probability  $|a_{m_1, m_2}(t)|^2$, of detecting a system in the state $|m_1\rangle   |m_2\rangle  $ is a continuous function; nonetheless, $\alpha$ for $\epsilon =0 $ is interrupted. Indeed, if $\lambda_k(\alpha)=\lambda_k(-\alpha)$, then $\alpha$ can be replaced by $\alpha\to |\alpha|$, where $|\alpha|=\sqrt{\epsilon^2+1}-|\epsilon|$, which is a continuous function. Unlike $\lambda_k$ and $|a_{m_1, m_2}(t)|^2$, the expression for the energy eigenvalues (\ref{10}) is not a definite function for $\epsilon =0 $. To determine the energy at the point $\epsilon =0 $, it is necessary to consider the Hamiltonian in the form ${\hat{H}}=\frac{1}{2}\left({\hat{H}}_{\epsilon \to 0+} +{\hat{H}}_{\epsilon \to 0-}\right) $. As a result, it is not difficult to obtain
\begin{eqnarray}
E_{n,m,\epsilon =0}=\frac{\omega_c}{2} \left( n+\frac{1}{2}\right)\left(  \sqrt{1+\frac{\beta \sqrt{\omega}}{\omega_c}}+\sqrt{1-\frac{\beta \sqrt{\omega}}{\omega_c}}\right) +
\nonumber\\ 
+\frac{\omega}{2}\left( m+\frac{1}{2}\right) \left(\sqrt{1+\frac{\beta \omega_c}{\omega^{3/2}}+\frac{\beta^2}{\omega}} + \sqrt{1-\frac{\beta \omega_c}{\omega^{3/2}}+\frac{\beta^2}{\omega}}\right) .
\label{23}
\end{eqnarray}

\begin{figure}[!hbp]
\begin{minipage}[h]{0.49\linewidth}
\center{\includegraphics[angle=0,width=1\textwidth, keepaspectratio]{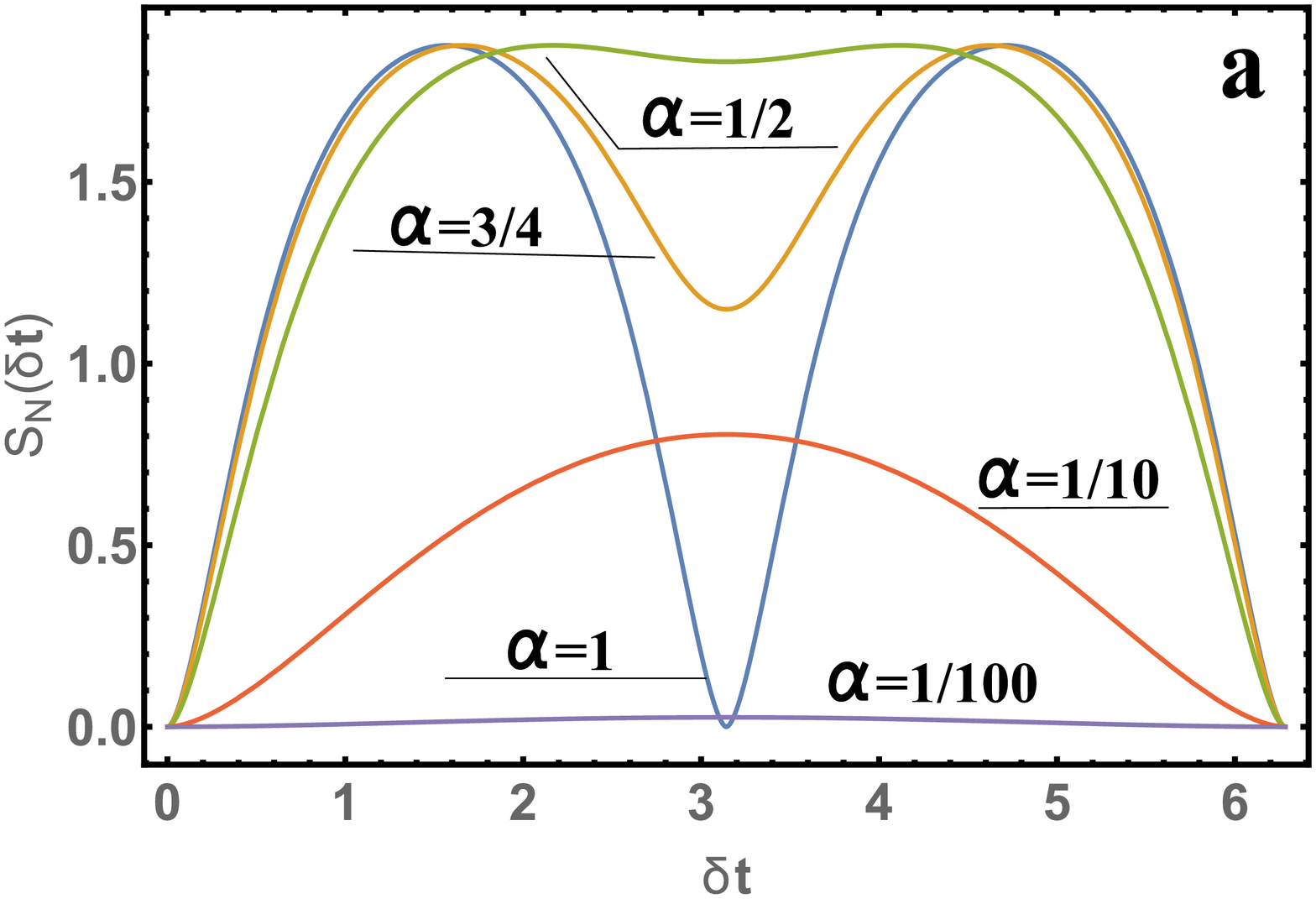}} \\
\end{minipage}
\hfill
\begin{minipage}[h]{0.49\linewidth}
\center{\includegraphics[angle=0,width=1\textwidth, keepaspectratio]{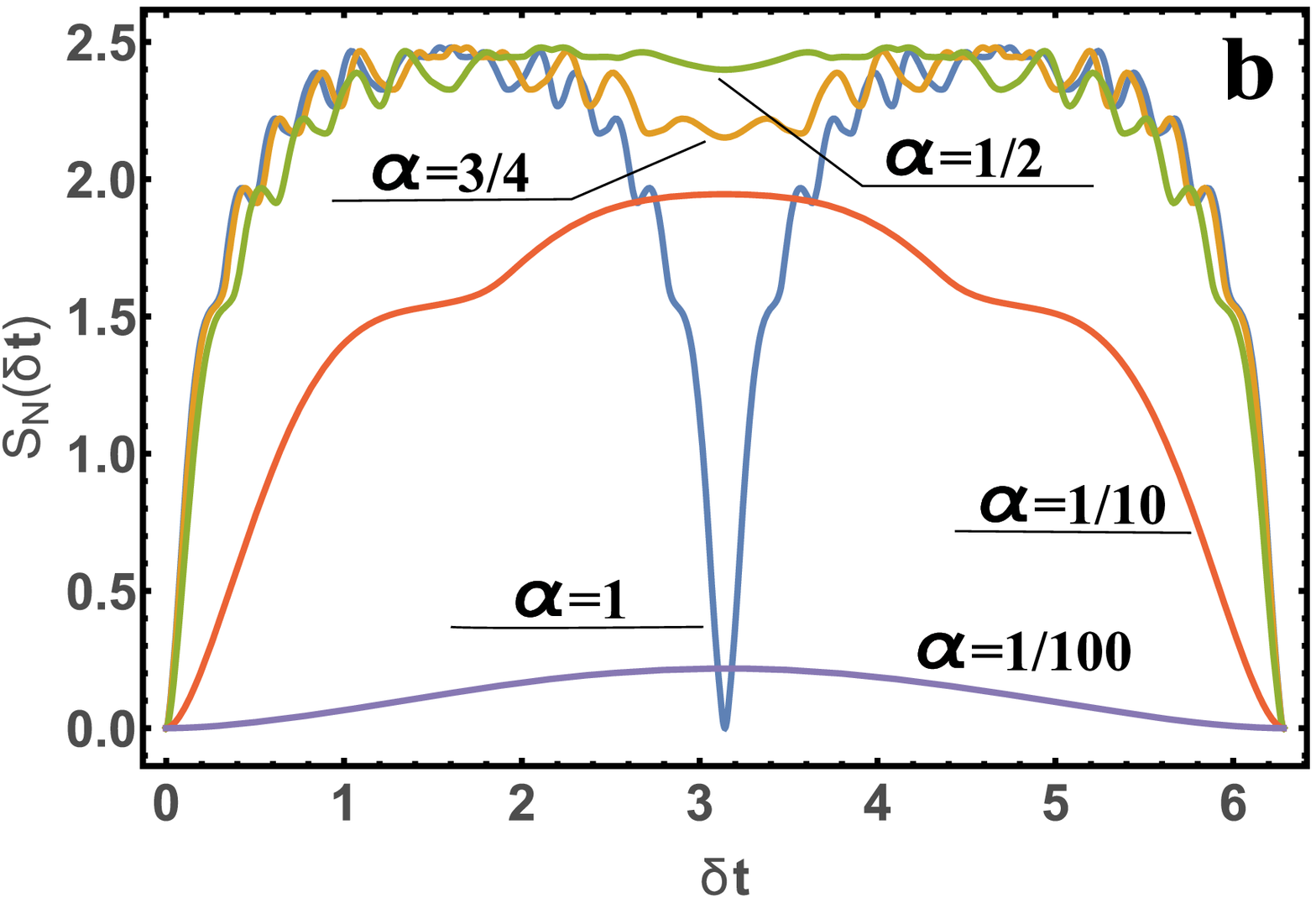}} \\ 
\end{minipage}
\hfill
\begin{minipage}[h]{0.49\linewidth}
\center{\includegraphics[angle=0,width=1\textwidth, keepaspectratio]{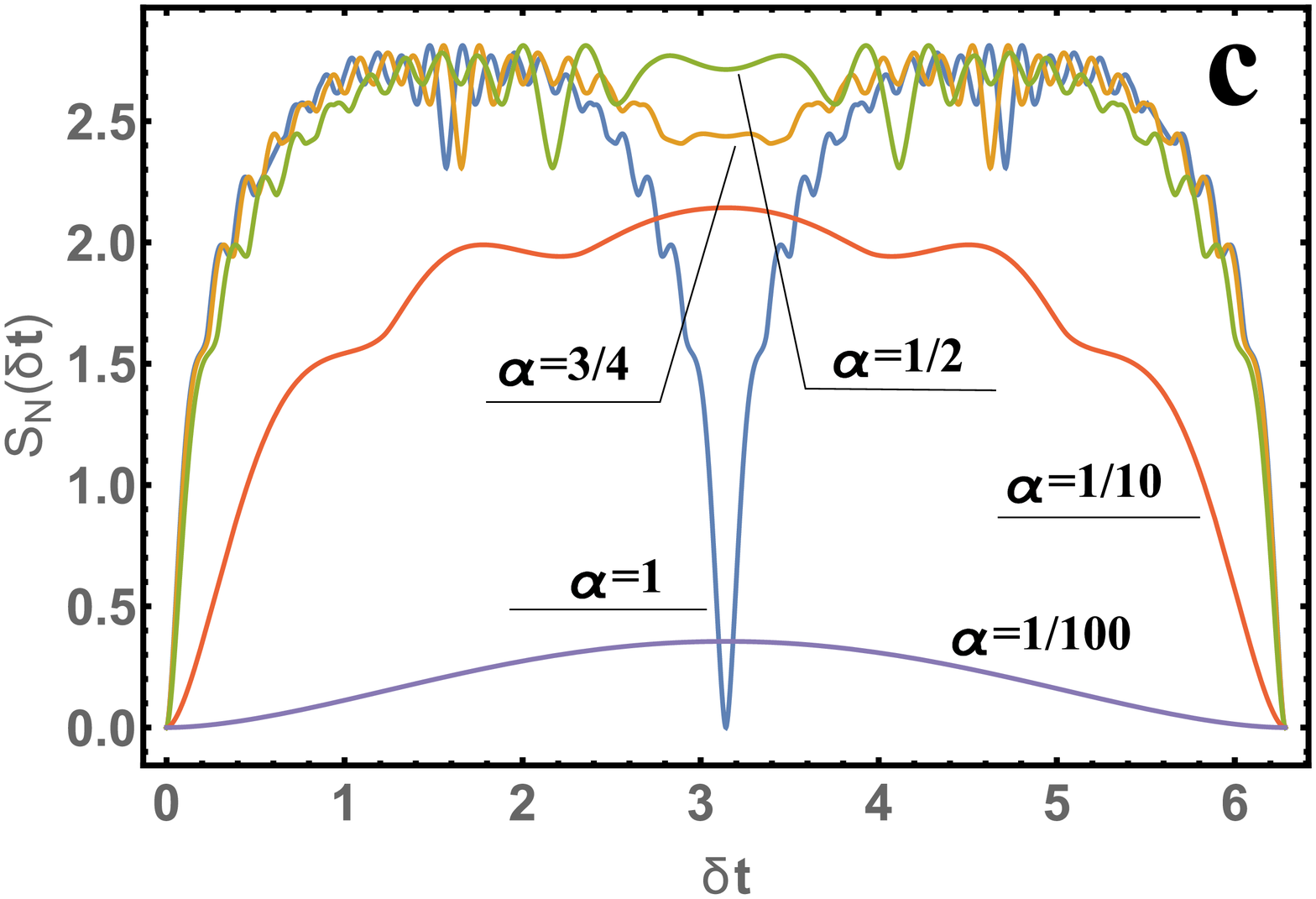}} \\
\end{minipage}
\hfill
\begin{minipage}[h]{0.49\linewidth}
\center{\includegraphics[angle=0,width=1\textwidth, keepaspectratio]{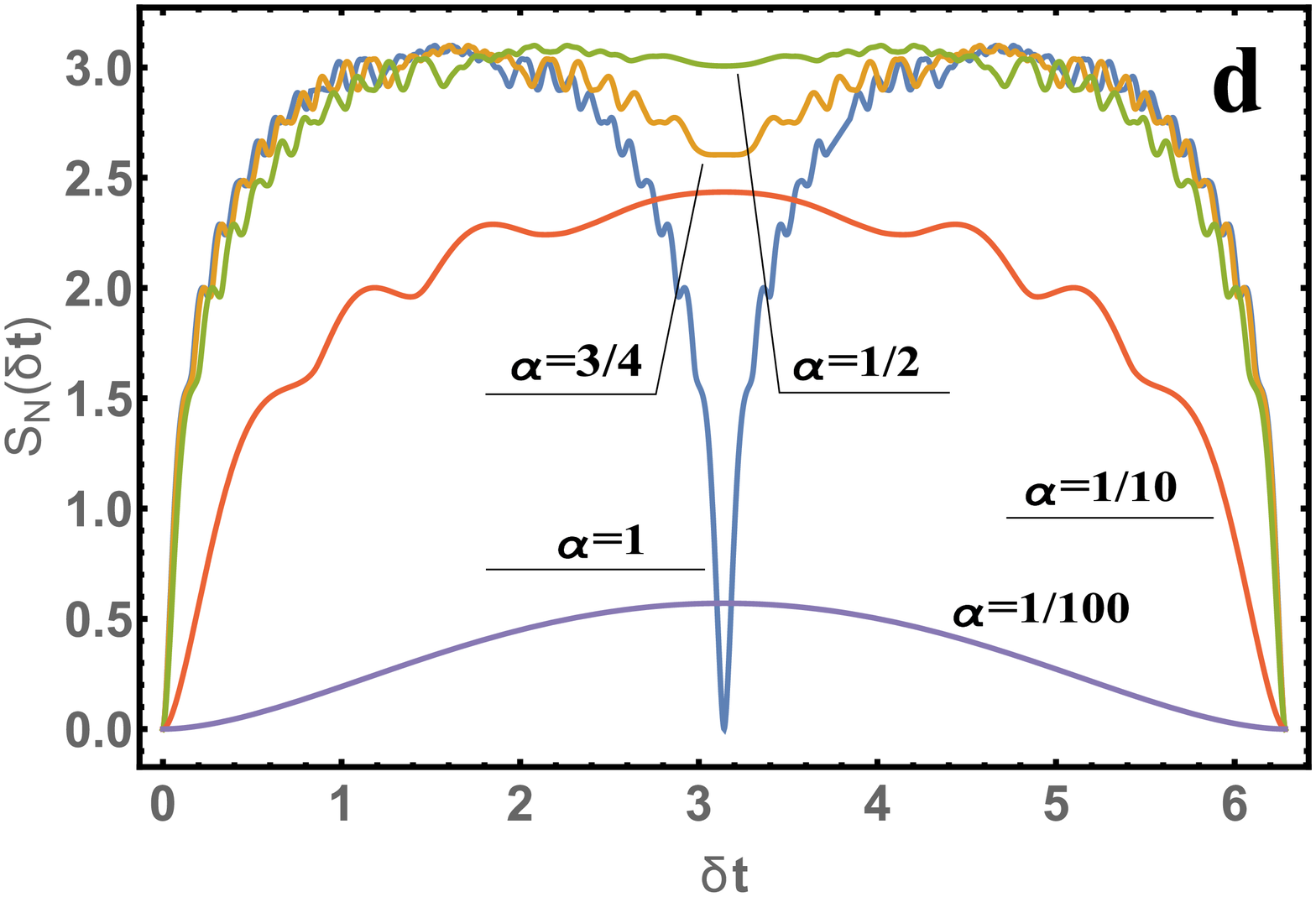}} \\
\end{minipage}
\center{\caption[fig]{ The results of calculating the Von Neumann entropy $S_N=S_N(\delta t)$ for  $\alpha = (1,3/4,1/2,1/10,1/100)$ and a) $s_1=0,s_2=10$; b) $s_1=5,s_2=10$; c) $s_1=10,s_2=10$; d) $s_1=20,s_2=10$  }}
\label{fig}
\end{figure}

\section{ Conclusion}

It can be seen from formula (\ref{20}) and Figure 1 that the entropy  $S_N=S_N(\delta t)$ is a $2\pi$ periodic function, i.e. $S_N(\delta t)=S_N(\delta t+2\pi)$.  It is also interesting to note that when $\delta \sim \beta$, the value is small; therefore, in order for the system to have a large quantum entanglement, it takes $t\sim 1/\beta$.  The system will quickly go into a quantum-entangled state in the case when $\omega_c$ is distinguishable from $\omega$, but when  $\beta<<1$ from expression (\ref{7}) it is clear that $\alpha<<1$ and $\gamma<<1$, so the quantum entanglement will be small. As a result, we can say that to achieve a large quantum entanglement of the system, $\omega_c \approx \omega$ is necessary, with $\Delta \omega =\omega_c - \omega$ less or about $\beta$, but the time for large quantum entanglement must be large $t\sim 1/\beta$.  In the opposite case  ($\omega_c$, as distinct from $\omega$), the system quickly becomes a quantum-entangled state, but the quantum entanglement is very small. Quantum entanglement will be a large value in the case of $s_1+s_2>>1$, for example, when the number of photons is large. It is known that the Von Neumann entropy has the maximum value $max\{S_N\}=\ln N$, where $N$ is the number of nonzero eigenvalues; in our case, therefore, $max\{S_N\}=\ln (s_1+s_2)$ (which is well illustrated by the figures). The advantage of the problem under investigation is that the quantum entanglement is not limited to the framework of the model and can take arbitrarily large values for a certain choice of $s_1$ and $s_2$.

In summary, a system consisting of a quantum harmonic oscillator interacting with a quantized electromagnetic field has been studied. The main result of this work is that it should now be possible to obtain in an analytic form the main characteristics of such a system as well as the probability of transitions and the Schmidt mode; quantum entanglement could also be studied further using, the Von Neumann entropy, and addition, such a system could also be used as an intensive source of quantum-entangled photons with an electron.

\end{document}